\documentclass[twoside, a4paper, nofootinbib,10pt]{article}
\usepackage{amssymb}
\usepackage{IEEEtrantools}
\usepackage[mathscr]{eucal}
\usepackage[dvips]{graphicx}
\usepackage{amsmath}
\usepackage{amsthm}
\usepackage{pstricks}
\usepackage{caption}
\usepackage{cite}
\usepackage{cancel}
\def \ni{\noindent}

\newcommand{\be}{\begin{equation}}
\newcommand{\ee}{\end{equation}}
\newcommand{\ben}{\begin{equation*}}
\newcommand{\een}{\end{equation*}}
\newcommand{\bes}{\begin{eqnarray}}
\newcommand{\ees}{\end{eqnarray}}
\newcommand{\besn}{\begin{IEEEeqnarray*}{rCl}}
\newcommand{\eesn}{\end{IEEEeqnarray*}}

\newcommand{\txt}{\textrm}

\newtheorem*{theorem*}{Theorem}

\newtheorem*{definition*}{Definition}
\newtheorem*{lemma*}{Lemma}
\newtheorem*{prop*}{Proposition}
\newtheorem*{corollary*}{Corollary}

\DeclareFontFamily{U}{mathx}{}
\DeclareFontShape{U}{mathx}{m}{n}{<-> mathx10}{}
\DeclareSymbolFont{mathx}{U}{mathx}{m}{n}
\DeclareMathAccent{\widecheck}{0}{mathx}{"71}

\title{Yang-Mills as a Constrained Gaussian}
\author{T. Tlas}
\date{}

\begin{document}
\maketitle
\thispagestyle{empty}

\begin{abstract}
\ni Yang-Mills is reformulated in terms of the logarithmic derivative of the holonomies. The classical equations of motion are recovered, and the path integral is rewritten in two ways, both of which are of the form of a Gaussian satisfying a quadratic constraint.
\end{abstract}

\section*{Introduction}

It is an old and well-known idea that the standard formulation of Yang-Mills may not be the most appropriate one to study its infrared, nonperturbative features. Therefore, there were many attempts to reformulate the theory in terms of other variables which would make these features more transparent. The most studied choice has been the Wilson loop, i.e. the trace of the holonomy of the connection. A large amount of effort was expended and numerous wonderful results were obtained (see e.g. chapter 12 in \cite{makeenko} and the references therein). Nonetheless, the reformulation in terms of Wilson loops was never fully completed. In particular the Yang-Mills path integral was never rewritten in terms of these variables. This is due to the complexity of the constraints which the Wilson loops should satisfy \cite{giles}.      \newline

Another set of variables, which, while known, has been rather under-studied, is the logarithmic derivative of the holonomy. It seems to have been first introduced by Mandelstam in \cite{mandelstam} and subsequently explored by several authors \cite{mandelstam2, mandelstam3, birula, polyakov, hong, hongbook, gross, gross2, driver}. The definitive analysis of the kinematics of these variables was performed by Gross in \cite{gross}, unifying and extending many of the previous results. We shall therefore refer below to these variables as Mandelstam-Gross variables. \newline

The aim of this manuscript is to supply a reformulation of Yang-Mills in terms of these variables both at the classical and quantum level. In the next section we shall rewrite the classical action in terms of them and then will recover the usual equation of motion.\footnote{The only attempt to do so seems to be in Appendix A of \cite{hong}. Our proof is different and we believe is more transparent.} In the section after that, we shall rewrite the Yang-Mills path integral in terms of these variables. This essentially amounts to computing the Jacobian of the transformation relating the Mandelstam-Gross variables to the connection.\newline

Before we proceed, we give a summary of our notation choices and a \textit{very} brief review of these variables. The reader at this stage should probably consult \cite{gross} as we will freely use the results from there.\newline

 We shall work in the Euclidean signature throughout. All the paths below are based at some arbitrary but fixed point. The parameter range for all the paths is $[0,1]$. Our gauge group will be $SU(N)$ throughout (though it is not difficult to generalize everything to an arbitrary simple Lie group). The Lie algebra of $SU(N)$ will be denoted by $su(N)$. Now, given an $SU(N)$ connection $A$ on $\mathbb{R}^D$, one can define $\mathcal{P}(\gamma)$ along any loop $\gamma$. Following the notation of \cite{gross}, the Mandelstam-Gross variables will be denoted by $\mathcal{B}_{\mu, s}$, where $\mu = 1, 2, \dots, D$ and $s \in (0,1)$, and are defined to be
 
 \ben
 \mathcal{B}_{\mu, s}(\gamma) = \Big (  \mathcal{P}(\gamma)  \Big)^{-1} \frac{\delta \mathcal{P}(\gamma)}{\delta \gamma(s)}.
 \een

We would like to emphasize that $(\mu, s)$ should be thought of as a single index, thus making $\mathcal{B}$ an $su(N)$-valued 1-form on the space of loops. \newline

It can be shown that 

\ben
\mathcal{B}_{\mu, s}(\gamma) = \Big ( \mathcal{P}(\gamma_s) \Big )^{-1} F_{\mu \nu}\big (\gamma(s) \big) \dot{\gamma}^\nu(s) \mathcal{P}(\gamma_s) \equiv \mathcal{F}_\mu (\gamma_s),
\een

where $F_{\mu \nu}$ stands for the curvature of the connection and $\gamma_s$ is the path $\gamma_s(t)  = \gamma(st)$.\newline

Given $C_{\mu, s}(\gamma)$, an $su(N)$-valued 1-form on the space of loops, we say that it is:

\begin{itemize}
\item[i-] transverse if $C_{\mu, s}(\gamma) \dot{\gamma}^\mu(s) = 0$ for every $s \in (0,1)$.
\item[ii-] nonanticipating if $\mathcal{C}_{\mu, s}(\gamma_1) = \mathcal{C}_{\mu, s}(\gamma_2)$ for all $\mu$ and all $s < s_0$ as long as $\gamma_1(s) = \gamma_2(s)$ for all $s < s_0$.
\end{itemize}

The crucial fact that was proven in \cite{gross} is that if an $su(N)$-valued 1-form is transverse, nonanticipating and has curvature zero, i.e. if

\ben
\mathcal{G}^{\mu, s; \nu, t}[\mathcal{C}] (\gamma) = \frac{\delta \mathcal{C}^{\nu,t}(\gamma)}{ \delta \gamma^\mu(s) } - \frac{ \delta \mathcal{C}^{\mu,s}(\gamma)}{\delta \gamma^\nu(t)} + \Big [ \mathcal{C}^{\mu,s}(\gamma), \mathcal{C}^{\nu,t}(\gamma)   \Big ] = 0,
\een

for all $\mu, \nu$ and for every $s, t \in (0,1)$, then there is a connection $A$ whose corresponding $\mathcal{B}$ is the given 1-form $\mathcal{C}_{\mu, s}(\gamma)$. Moreover, one can actually recover $A$ from $\mathcal{B}$ by a linear map $\mathcal{T}$ given by

\be
\label{eq:tmap}
A_\mu(x) = \mathcal{T}(\mathcal{B}) \equiv \int_0^{\frac{1}{2}} ds \,  \mathcal{B}_{\mu, s}(\sigma_x) \partial^\mu \sigma_x(s),
\ee

where $\sigma_x$ is the loop which is given by

\ben
\sigma_x(s) = \begin{cases}
2 s x & , \quad 0 \leq s \leq \frac{1}{2}  \\
2(1-s) x & , \quad \frac{1}{2} \leq s \leq 1.
\end{cases}
\een

It can be shown that the connection $\mathcal{T}(\mathcal{B})$ satisfies the radial gauge, and that the map $\mathcal{T}$ establishes a bijection between the set of all transverse, nonanticipating, curvature zero 1-forms on the space of loops and the set of all connections satisfying the radial gauge with the composition $A \to \mathcal{B}(A) \to \mathcal{T}(\mathcal{B}(A))$ being the identity on the space of such connections. It is assumed below that wherever gauge-fixing is needed, the radial gauge is used.

\section*{The Classical Action and the Equations of Motion}

We begin by giving the Yang-Mills action in terms of the Mandelstam-Gross variables \cite{hong}.  It is given by the following formula

\be
\label{eq:action}
\frac{D}{\mathcal{V}} \int_{(0,1)} ds \int \mathcal{D} \gamma \frac{  \txt{Tr}   \Big ( \mathcal{B}_{\mu, s}(\gamma) \mathcal{B}^{\mu, s}(\gamma)  \Big )   }{ \dot{\gamma}_\nu(s) \dot{\gamma}^\nu(s)   },
\ee

where $D$ is the dimensions of the spacetime and $\mathcal{D} \gamma$ and $\mathcal{V}$ stand for the following formal expressions

\ben
\mathcal{D} \gamma  =  \prod_{t \in (0,1)} d \gamma(t),  \quad \txt{and} \quad   \mathcal{V}  =  \int \prod_{t \in (0,1); t \neq s} d \gamma(t).
\een

In other words, $\mathcal{D} \gamma$ stands for the (nonexistent) `flat' measure on the space of loops $\Gamma$, and $\mathcal{V}$ is the (infinite) volume assigned by the measure to a `slice' of the space, obtained from $\mathcal{D} \gamma$ by peeling off the integration over $\gamma(s)$. \newline

Of course, at face value, these expressions are ill-defined. Nonetheless, we are going to continue using them below for two reasons. First, it simplifies the formal calculations considerably, mainly because it allows integration by parts. Second, it is in fact rather easy, if desired, to make sense of the formal expressions above by interpreting $\mathcal{D} \gamma$ to stand for an appropriate, well-defined measure. What measure would classify as appropriate? As we shall see below, it should satisfy the following properties:

\begin{itemize}
\item[i-] $\mathcal{D} \gamma_x$ is supported on loops which are at least $C^1$.
\item[ii-] $\int \mathcal{D} \gamma_x  < \infty$.
\item[iii-] For any $s \in (0,1)$, the integral $\int \mathcal{D} \gamma$ should factorize into an integral over the value of $\gamma$ and $\dot{\gamma}$ at $s$, an integral over the part of $\gamma$ from 0 to $s$ and a final integral over the part from $s$ to 1.
\item[iv-] The measure corresponding to the integral over $\dot{\gamma}(s)$ in the previous item is even.
\end{itemize}

Contrary to what it might seem, it is rather straightforward to construct measures which satisfy the conditions above. In fact the Brownian bridge measure almost fits the bill. The only issue is that the Brownian paths are not sufficiently regular. An obvious substitute would be a Gaussian measure whose covariance is sufficiently smoothing. The simplest choice for a covariance that works is to take the bilinear form $\frac{1}{(- \nabla^2 + 1)^2}$. An additional attractive feature is that one can have an entire family of measures $\mathcal{D} \gamma_\epsilon$ parametrized by $\epsilon > 0$, whose covariances are given by $ \frac{1}{\epsilon ( - \nabla^2 + 1)^2}$, such that as $\epsilon \to 0$, they can be thought to approach the `flat' one. This would justify the integration by parts manipulations in what follows. \newline

Now, let us demonstrate that (\ref{eq:action}) is indeed the Yang-Mills action in disguise. Note that

\besn
\frac{D}{\mathcal{V}} \int_{(0,1)} ds \int \mathcal{D} \gamma \frac{  \txt{Tr}   \Big ( \mathcal{B}_{\mu, s}(\gamma) \mathcal{B}^{\mu, s}(\gamma)  \Big )   }{ \dot{\gamma}_\nu(s) \dot{\gamma}^\nu(s)   } & = & \\
\frac{D}{\mathcal{V}} \int_{(0,1)} ds  \int \mathcal{D} \gamma \frac{  \txt{Tr}   \Big ( \mathcal{F}_\mu(\gamma_s) \mathcal{F}^\mu(\gamma_s)  \Big )   }{ \dot{\gamma}_\nu(s) \dot{\gamma}^\nu(s)   } &  = & \\
\frac{D}{\mathcal{V}} \int_{(0,1)} ds \int d\gamma(s) \int \prod_{t \in (0,1); t \neq s} d \gamma(t) \frac{ \txt{Tr} \Big ( F_{\mu \rho}(\gamma(s)) F^{\mu \rho'}(\gamma(s))  \Big ) \dot{\gamma}^\rho(s) \dot{\gamma}_{\rho'}(s)  }{ \dot{\gamma}_{\nu}(s) \dot{ \gamma}^\nu(s)  }  & = & \\
 \int_{(0,1)} ds \int dx \,  \txt{Tr} \Big ( F_{\mu \rho}(x) F^{\mu \rho'}(x)   \Big ) \delta^\rho_{\rho'}   & = & \\
 \int dx \, \txt{Tr} \Big ( F_{\mu \rho}(x) F^{\mu \rho}(x)   \Big ).
\eesn

The above calculation carries through if we use a well-defined measure satisfying (i)-(iv). The first two items will make the expression for the action well-defined. The factorization described in (iii) would be the analogue of the factorization in the third line above, except that we will have an additional integration over $\dot{\gamma}(s)$ (here it is subsumed in the $\prod_{t \in (0,1); t \neq s} d \gamma(t)$). Finally, the last item would reproduce the Kronecker delta in the fourth line.\newline

We want to show now that we recover the usual Yang-Mills equations of motion by setting to zero the variation of the action (\ref{eq:action}) over the set of transverse, nonanticipating $\mathcal{B}_{\mu,s}$'s satisfying the constraint 

\be
\label{eq:constraint}
\mathcal{G}^{\mu, s; \nu, t} (\gamma) = \frac{\delta \mathcal{B}^{\nu,t}(\gamma)}{ \delta \gamma^\mu(s) } - \frac{ \delta \mathcal{B}^{\mu,s}(\gamma)}{\delta \gamma^\nu(t)} + \Big [ \mathcal{B}^{\mu,s}(\gamma), \mathcal{B}^{\nu,t}(\gamma)   \Big ] = 0.
\ee

At this point, one can use the result (see Theorem 3.13 and Corollary 3.5 in \cite{gross}) that (under suitable smoothness assumptions) $\mathcal{B}^{\mu,s}$'s provide a faithful coordinatization of the restricted gauge orbit space. Since the action is a smooth function(al) on this space, and since the critical points of a function(a) are unaffected by changes of variables applied to its domain, we have what we want. Despite the fact that there is nothing wrong with the argument above, it would certainly be more satisfying if one can supply a more explicit computational proof.\newline

Therefore, in order to perform the variation of (\ref{eq:action}) subject to the constraint (\ref{eq:constraint}), we introduce the Lie algebra valued Lagrange multiplier $\Xi_{\mu,s ; \nu, t}(\gamma)$. Since $\mathcal{G}^{\mu, s; \nu, t}$ is antisymmetric under $\mu, s \longleftrightarrow \nu, t$, we can, without loss of generality, assume that $\Xi_{\mu, s; \nu, t}$ is antisymmetric as well. We thus add the following term 

\be
\label{eq:caction}
\int \mathcal{D}\gamma \int_{(0,1)^2} ds \, dt  \, \txt{Tr} \Big (  \Xi_{\mu, s; \nu, t}(\gamma) \, \mathcal{G}^{\mu, s; \nu, t} (\gamma) \Big )
\ee

to our action and vary $\mathcal{B}^{\mu,s}$ and $\Xi_{\mu, s; \nu, t}$. Of course, variation of the $\Xi$ simply gives back the constraint. We therefore only need to vary the $\mathcal{B}$'s. Performing the variation is straightforward as the full action is quadratic in the $\mathcal{B}$'s with the result being the following equation

\be
\label{eq:var}
\frac{ \mathcal{B}^{\mu,s}(\gamma)   }{ \mathcal{V} \dot{\gamma}_\rho(s) \dot{\gamma}^\rho(s)    } =   \int_{(0,1)}dt \bigg (   \frac{ \delta \Xi_{\mu, s; \nu, t}(\gamma)  }{  \delta \gamma^\nu(t)  } +  \Big [  \mathcal{B}^{\nu,t}(\gamma)  ,   \Xi_{\mu, s; \nu, t}(\gamma)  \Big ] \bigg ).
\ee

Before we proceed, let us identify our target, i.e. let us give the Yang-Mills equations of motion in terms of the Mandelstam-Gross variables. This is well-known \cite{polyakovbook} and is given by the expression

\ben
\frac{ \delta \mathcal{F}^\mu(\gamma')  }{ \delta \gamma^{'\mu}(1)    } = 0,
\een

where this equation should hold for any path $\gamma'$. \newline

Now, note that 

\besn
\int_{(0,1)} ds \Bigg (  \frac{\delta }{ \delta \gamma^{\mu}(s)  } \bigg ( \frac{ \mathcal{B}^{\mu,s}(\gamma)   }{ \mathcal{V} \dot{\gamma}_\rho(s) \dot{\gamma}^\rho(s)    }  \bigg ) \Bigg )& = & \\
\int_{(0,1)} ds \Bigg (  \frac{\delta }{ \delta \gamma^{\mu}(s)  }   \bigg ( \frac{ \mathcal{F}^{\mu}(\gamma_s)   }{ \mathcal{V} \dot{\gamma}_\rho(s) \dot{\gamma}^\rho(s)    }  \bigg ) \Bigg ) &= & \\
\int_{(0,1)} ds \Bigg (   \frac{\frac{\delta  \mathcal{F}^\mu(\gamma_s)  }{ \delta \gamma^{\mu}(s)  } }{ \mathcal{V} \dot{\gamma}_\rho(s) \dot{\gamma}^\rho(s)   } -  \mathcal{F}^\mu(\gamma_s) \frac{  2 \dot{\delta}(0) \delta^\rho_\mu \dot{\gamma}_\rho(s)  }{  \mathcal{V} \big ( \dot{\gamma}_\rho(s) \dot{\gamma}^\rho(s) \big )^2  } \Bigg ) &  = & \\
\int_{(0,1)} ds \Bigg (  \frac{1}{  \mathcal{V} \dot{\gamma}_\rho(s) \dot{\gamma}^\rho(s)  } \frac{\delta  \mathcal{F}^\mu(\gamma_s)  }{ \delta \gamma_s ^{\mu}(1)  } \Bigg ),
\eesn

where the second term on the third line vanishes (no matter how we define $\dot{\delta}(0)$) in view of the transversality constraint that the $\mathcal{B}$'s (and hence the $\mathcal{F}$'s) satisfy. Clearly, we have that if $ \frac{ \delta \mathcal{F}^\mu(\gamma')  }{ \delta \gamma^{'\mu} (1)    } = 0$ for every path $\gamma'$, then $$ \int_{(0,1)} ds \Bigg (  \frac{\delta }{ \delta \gamma^{\mu}(s)  } \bigg ( \frac{ \mathcal{B}^{\mu,s}(\gamma)   }{ \mathcal{V} \dot{\gamma}_\rho(s) \dot{\gamma}^\rho(s)    }  \bigg ) \Bigg ) = 0$$ for every loop $\gamma$. We claim that the converse implication holds as well. To see this, let $\gamma$ be a loop and let $s_0 \in (0,1)$. We know that

\besn
0 & = & \int_{(0,1)} ds \Bigg (  \frac{1}{  \mathcal{V} \dot{\gamma}_\rho(s) \dot{\gamma}^\rho(s)  } \frac{\delta  \mathcal{F}^\mu(\gamma_s)  }{ \delta \gamma_s ^{\mu}(1)  } \Bigg )   \\
& = & \int_0^{s_0} ds \Bigg (  \frac{1}{  \mathcal{V} \dot{\gamma}_\rho(s) \dot{\gamma}^\rho(s)  } \frac{\delta  \mathcal{F}^\mu(\gamma_s)  }{ \delta \gamma_s ^{\mu}(1)  } \Bigg ) +  \int^1_{s_0} ds \Bigg (  \frac{1}{  \mathcal{V} \dot{\gamma}_\rho(s) \dot{\gamma}^\rho(s)  } \frac{\delta  \mathcal{F}^\mu(\gamma_s)  }{ \delta \gamma_s ^{\mu}(1)  } \Bigg ).
\eesn

If we replace $\gamma$ with another loop $\tilde{\gamma}$ which coincides with $\gamma$ for $t \leq s_0$, then the first term above does not change. The second term can be made as small as one pleases by having the $\dot{\tilde{\gamma}}(t)$ very large while $\tilde{\gamma}(t)$ stays bounded for $t \geq s_0$.\footnote{An objection might be raised here that there is a hidden factor of the $\dot{\tilde{\gamma}}$ hidden inside $\mathcal{F}^\mu$. This is indeed the case, but the denominator in $ \int^1_{s_0} ds \Bigg (  \frac{1}{  \mathcal{V} \dot{\gamma}_\rho(s) \dot{\gamma}^\rho(s)  } \frac{\delta  \mathcal{F}^\mu(\gamma_s)  }{ \delta \gamma_s ^{\mu}(1)  } \Bigg )$ has two such factors and thus dominates.} Therefore, we can conclude that for any $s_0 \in (0,1)$ 

\ben
 \int_0^{s_0} ds \Bigg (  \frac{1}{  \mathcal{V} \dot{\gamma}_\rho(s) \dot{\gamma}^\rho(s)  } \frac{\delta  \mathcal{F}^\mu(\gamma_s)  }{ \delta \gamma_s ^{\mu}(1)  } \Bigg ) = 0.
\een

From this it follows that the integrand itself vanishes and we have our claim.\newline

The upshot of the discussion above is that, to verify that the Yang-Mills equations of motion hold, it is sufficient to apply $\int_{(0,1)} ds \frac{\delta }{ \delta \gamma^{\mu}(s)  }$ to the right hand side of (\ref{eq:var}) and then verify that the result vanishes. \newline

Therefore consider

\be
\label{eq:divergence}
 \int _{(0,1)^2} ds \, dt \bigg (  \frac{ \delta^2 \Xi_{\mu, s; \nu, t}(\gamma)  }{ \delta \gamma^\mu(s) \delta \gamma^\nu(t)  } +  \Big [  \frac{\delta \mathcal{B}^{\nu,t}(\gamma)}{ \delta \gamma^\mu(s)    }  ,   \Xi_{\mu, s; \nu, t}(\gamma)  \Big ] +  \Big [  \mathcal{B}^{\nu,t}(\gamma)  ,   \frac{  \delta \Xi_{\mu, s; \nu, t}(\gamma)}{ \delta \gamma^\mu(s)    }  \Big ] \bigg ) .
\ee

The first term vanishes due to the antisymmetry of $\Xi$. The second term can be rewritten in the following way

\bes
\nonumber
 \int _{(0,1)^2} ds \, dt \bigg (  \Big [  \frac{\delta \mathcal{B}^{\nu,t}(\gamma)}{ \delta \gamma^\mu(s)    }  ,   \Xi_{\mu, s; \nu, t}(\gamma)  \Big ]    \bigg ) & = & \\
 \nonumber
   \int _{(0,1)^2} ds \, dt \bigg (  \frac{1}{2} \Big [  \frac{\delta \mathcal{B}^{\nu,t}(\gamma)}{ \delta \gamma^\mu(s)    } - \frac{\delta \mathcal{B}^{\mu,s}(\gamma)}{ \delta \gamma^\nu(t)    }   ,   \Xi_{\mu, s; \nu, t}(\gamma)  \Big ] \bigg ) & = &  \\
  \label{eqs:first}
  - \int _{(0,1)^2} ds \, dt\bigg ( \frac{1}{2}  \Big [  \big [  \mathcal{B}^{\mu, s}(\gamma) , \mathcal{B}^{\nu, t}(\gamma)    \big ]  ,   \Xi_{\mu, s; \nu, t}(\gamma)  \Big ] \bigg ),
\ees

where the antisymmetry of $\Xi$ justifies the first equality, and (\ref{eq:constraint}) justifies the second. \newline

Looking at the third term in (\ref{eq:divergence}), and using (\ref{eq:var}) we get

\bes
\nonumber
 \int _{(0,1)^2} ds \,  dt \bigg ( \Big [  \mathcal{B}^{\nu,t}(\gamma)  ,   \frac{  \delta \Xi_{\mu, s; \nu, t}(\gamma)}{ \delta \gamma^\mu(s)    }  \Big ] \bigg ) & = & \\
 \nonumber
\int _{(0,1)^2} ds \,  dt \bigg (   \Big [  \mathcal{B}^{\nu,t}(\gamma)  , -  \frac{ \mathcal{B}^{\nu,t}(\gamma)   }{ \mathcal{V} \dot{\gamma}_\rho(t) \dot{\gamma}^\rho(t)    }  + \Big [  \mathcal{B}^{\mu,s}(\gamma)  ,   \Xi_{\nu, t; \mu, s}(\gamma)  \Big ]         \bigg )  & = & \\
 \label{eqs:second}
  \int _{(0,1)^2} ds \,  dt \bigg (   \Big [  \mathcal{B}^{\nu,t}(\gamma)  ,  \Big [  \mathcal{B}^{\mu,s}(\gamma)  ,   \Xi_{\nu, t; \mu, s}(\gamma)  \Big ]         \bigg ) .
\ees

Now, adding together (\ref{eqs:first}) and (\ref{eqs:second}) and dropping the $\gamma$'s to reduce clutter we get

\besn
 \int _{(0,1)^2} ds \, dt\bigg \{ - \frac{1}{2}  \Big [  \big [  \mathcal{B}^{\mu, s} , \mathcal{B}^{\nu, t}    \big ]  ,   \Xi_{\mu, s; \nu, t}  \Big ] +  \Big [  \mathcal{B}^{\nu,t}  ,  \Big [  \mathcal{B}^{\mu,s} ,   \Xi_{\nu, t; \mu, s} \Big ]         \bigg \}& = & \\
 \int_{(0,1)^2} ds \, dt \bigg \{   - \frac{1}{2} \Big ( \mathcal{B}^{\mu,s} \mathcal{B}^{\nu,t} \Xi_{\mu, s; \nu, t}   - \mathcal{B}^{\nu, t} \mathcal{B}^{\mu,s} \Xi_{\mu, s; \nu, t} - \Xi_{\mu, s; \nu, t} \mathcal{B}^{\mu, s} \mathcal{B}^{\nu, t} & + & \\ \Xi_{\mu,s; \nu, t} \mathcal{B}^{\nu, t} \mathcal{B}^{\mu, s} ) \Big )  + \Big (  \mathcal{B}^{\nu,t} \mathcal{B}^{\mu,s} \Xi_{\nu, t; \mu, s}  - \cancel{\mathcal{B}^{\nu, t} \Xi_{\nu, t; \mu, s} \mathcal{B}^{\mu, s}} - \cancel{\mathcal{B}^{\mu, s} \Xi_{\nu, t; \mu, s} \mathcal{B}^{\nu, t} } & + & \\
 \Xi_{\nu, t; \mu, s} \mathcal{B}^{\mu, s} \mathcal{B}^{\nu, t} \Big ) \bigg \} & = & \\
 \int _{(0,1)^2} ds \, dt \bigg (  \Big (      \mathcal{B}^{\nu, t} \mathcal{B}^{\mu, s} \Xi_{\mu, s; \nu, t} + \Xi_{\mu, s; \nu, t} \mathcal{B}^{\mu, s} \mathcal{B}^{\nu, t}           \Big ) & + &  \\
 \Big (  \mathcal{B}^{\nu, t} \mathcal{B}^{\mu, s} \Xi_{\nu, t; \mu, s}  + \Xi_{\nu, t; \mu, s} \mathcal{B}^{\mu, s} \mathcal{B}^{\nu, t}              \Big )   \bigg )  =  & 0 &.
\eesn 

We have thus recovered the Yang-Mills equations of motion.

\section*{The Path Integral}

Our goal now is to extend the reformulation of Yang-Mills in terms of the $\mathcal{B}$'s to the quantum setting. We shall do this by rewriting the Yang-Mills path integral in terms of the new variables. As we've mentioned above, this amounts to computing the Jacobian of the change of variables $A \to \mathcal{B}(A)$. This shall be done in three steps. First, we'll derive the relevant finite dimensional formulae. This will be the longest part of the calculation. Then we'll calculate the Jacobian for the closely related case of the principal chiral model. We will finish by dealing with the Yang-Mills case.

\subsection*{Finite Dimensions}

Suppose that we have the following three maps

\besn
F & : & \mathbb{R}^{n+m}   \to  \mathbb{R}  \qquad \, \,\,\, \txt{``the integrand''} \\
g & :  & \mathbb{R}^n  \to  \mathbb{R}^{n+m} \qquad   \, \txt{``the parametrization"} \\
h & : & \mathbb{R}^{n+m}  \to  \mathbb{R}^l \qquad    \txt{``the constraint''}
\eesn

We are going to suppose that all the maps above are as smooth as is required. Moreover, we will assume that $g$ is an embedding, i.e. that it is an immersion and that $g(\mathbb{R}^n)$ is an embedded $n$-dimensional submanifold of $\mathbb{R}^{n+m}$. We shall denote this submanifold by $S$. Finally, we will assume that $g(\mathbb{R}^n) = S = h^{-1}(0)$. \newline

We are interested in rewriting 

\ben
\int_{\mathbb{R}^n} F(g(x)) dx
\een

in terms of an integral over $\mathbb{R}^{n+m}$ with an insertion of a delta function containing $h$. As a first step, we rewrite the integral above in terms of the Hausdorff measure on the submanifold $S$. Then, using the area formula\footnote{See e.g. Theorem 3.9 in \cite{evans}.} we have the following equality

\be
\label{eq:param}
\int_{\mathbb{R}^n} F(g(x)) dx = \int_S F(y) \frac{1}{Jg(y)}d \mathcal{H}(y),
\ee

where $d \mathcal{H}$ stands for the Hausdorff measure on $S$. There are two equivalent definitions of the Jacobian factor $Jg(y)$.\footnote{There are slightly different conventions for the normalization of the Hausdorff measure in the literature, and formula (\ref{eq:param}) above holds for only one of them. The exact choice of convention however is not going to matter since below, we will only care about the proportionality of two expressions and not their equality.} The more familiar one is

\be
\label{eq:jacobianparam}
Jg(y) = \sqrt{   \det \big (  (Dg^\dagger) \big \vert_{g^{-1}(y)} Dg \big \vert_{g^{-1}(y)}  \big )   },
\ee

where $Dg$ is the derivative of $g$. The second definition \cite{krantz} of $Jg(y)$ is

\be
\label{eq:jacobian}
Jg(y) = \txt{volume of the parallelepiped spanned by } Dg(v_1), \dots, Dg(v_n),
\ee

where $Dg$ is evaluated at $g^{-1}(y)$ and $v_1, \dots, v_n$ are an orthonormal basis in $\mathbb{R}^n$. The word `volume' is understood to be the $n$-dimensional Hausdorff measure. \newline

Formula (\ref{eq:param}) can be easily generalized to the case when the domain of the ``parametrization'' $g$ instead of being $\mathbb{R}^n$ is an $n$-dimensional Riemannian manifold $M$. In this case, (\ref{eq:param}) is modified to become

\be
\label{eq:gparam}
\int_M F(g(x)) d \txt{vol}(x) = \int_S F(y) \frac{1}{Jg(y)}d \mathcal{H}(y),
\ee

where $d \txt{vol}$ stands for the volume form induced from the Riemannian metric on $M$. $Jg$ is again given by (\ref{eq:jacobian}) where $v_1, \dots, v_n$ are now an orthonormal basis of the tangent plane to a point of $M$.\footnote{The analogue of (\ref{eq:jacobianparam}) also holds but we won't need it.} \newline

Now, let $K: \mathbb{R}^{n+m} \to \mathbb{R}$ be a function and assume for now that $h$ is a submersion (and thus $l = m$). Using the coarea formula\footnote{See e.g. Theorem 3.11 in \cite{evans}.}, we have that

\be
\label{eq:coarea}
\int_{\mathbb{R}^{n+m}} K(z) \frac{e^{ - \frac{ h^2(z) }{\epsilon}  }    }{ (\sqrt{ \pi \epsilon})^m} Jh(z) dz = \int_{\mathbb{R}^m} \frac{  e^{- \frac{c^2}{\epsilon}   }  }{ (\sqrt{  \pi \epsilon } )^m   } \bigg ( \int_{ h^{-1}(c) }  K(t) d \mathcal{H}(t)     \bigg ) dc.
\ee

Again, analogously to $Jg(y)$, there are two definitions for $Jh(z)$. The first one is the expression

\be
\label{eq:jacobianlevel}
Jh(z) = \sqrt{\det \Big ( Dh\big  \vert_z (Dh)^\dagger \big \vert_z   \Big )          }.
\ee

The second one, which will play a much bigger role below, is 

\be
\label{eq:jacobian1}
Jh(z) = \txt{volume of the parallelepiped spanned by } Dh(v_1), \dots, Dh(v_l),
\ee

where $v_1, \dots, v_l$ is an othonormal basis of the orthogonal complement of the kernel of of $Dh$.\newline

If we now send $\epsilon \to 0$, let $K(z) = F(z) \frac{1}{Jg(z)}$, and recall that $S = h^{-1}(0)$, we get that

\ben
\int_S F(y) \frac{1}{Jg(y)}d \mathcal{H}(y) = \int_{\mathbb{R}^{n+m}} F(z) \delta \big (h(z) \big ) \frac{Jh(z)}{Jg(z)} dz.
\een

We would like to generalize this expression to the case when $h$ is \textit{not} a submersion, but is of constant rank. Therefore, suppose that the rank of $h$ is $m \leq l$. Assume first that, in fact, $Dh$ (and thus $h$) maps $\mathbb{R}^{n+m}$ into the $m$-dimensional subspace of $\mathbb{R}^l$ consisting of all those $l$-tuples whose last $l-m$ entries are zero. Formula (\ref{eq:coarea}) still applies, but there are two important differences. First, only definition (\ref{eq:jacobian1}) of $Jh$ works as long as `volume' is understood to mean the $m$-dimensional Hausdorff measure in $\mathbb{R}^l$. The second is that the limit $\epsilon \to 0$ does not produce $\delta(h)$ on the left hand side. If we denote the components of $h$ by $(h_1, h_2, \dots, h_l)$, then \textit{very}\footnote{$\delta(h)$ is in fact badly defined, since it contains factors of the form $\delta(0)$. This is less of a problem than it seems for two reasons. First, as discussed below, constant factors (even if formally infinite) can often be safely ignored in path integral calculations. Second, the delta functions are formal expressions which are usually implemented through a limiting procedure (or a Fourier transform), and the intermediate expressions are perfectly well-defined with the only problem arising, like here, at the limit (see the discussion about the `soft' imposition of constraints in \cite{construction}).} formally,

\be
\label{eq:delta}
\delta(h) = \delta(h_1) \delta(h_2) \dots \delta(h_l) = \lim_{\epsilon \to 0} \frac{1}{(\sqrt{\pi \epsilon} )^l } e^{- \frac{h_1^2}{\epsilon}} \dots e^{- \frac{h_l^2}{\epsilon}}  = \lim_{\epsilon \to 0}  \frac{e^{ - \frac{ h^2(z) }{\epsilon}  }    }{ (\sqrt{ \pi \epsilon})^l}   ,
\ee

which differs by a factor $\sqrt{\pi \epsilon}^{l-m}$ from the expression appearing in (\ref{eq:coarea}). \newline

Note however that our end goal is to compute certain path integrals, and in such computations it is not the value of the integral that matters, but its logarithmic derivatives with respect to the sources. Therefore, introducing the notation $\simeq$ which stands for ``proportional"\footnote{Of course, any two things are proportional. The nontrivial statement in our case is that the same constant makes two integrals equal \textit{for all} $F$'s.} and keeping in mind that $\delta(h)$ should be understood as a limit (or perhaps through its Fourier transform) we can write that

\be
\label{eq:glevel}
\int_{\mathbb{R}^{n+m}} K(z) \delta(h(z)) Jh(z) dz \simeq \int_S K(y) d \mathcal{H}(y).
\ee

Above, we've assumed that the range of $Dh(z)$ is a certain fixed $\mathbb{R}^m$ for all $z$. There is no difficulty however in generalizing the arguments above to the general case. Simply replace $h(z)$ with $\tilde{h}(z) = O(z) h(z)$ where $O(z)$ it he rotation which takes $Dh(z)$ to the $m$-dimensional subspace consisting of all those $l$-tuples whose last $l-m$ entries vanish. All the arguments above then apply, and equation (\ref{eq:glevel}) applies with $h$ replaced by $\tilde{h}$. However, as is evident from (\ref{eq:jacobian1}) and (\ref{eq:delta}), $Jh(z) = J\tilde{h}(z)$ and $\delta(h) = \delta(\tilde{h})$. Therefore, we can see that (\ref{eq:glevel}) holds for any $h$ of constant rank. \newline

Now, combining (\ref{eq:gparam}) and (\ref{eq:glevel}), and setting $K(z) = F(z) \frac{1}{Jg(z)}$, we get that

\be
\label{eq:relation}
\int_M F(g(x)) d\txt{vol}(x) \simeq \int_{\mathbb{R}^{n+m}} F(z) \delta \big (h(z) \big ) \frac{Jh(z)}{Jg(z)} dz. \ee

\vspace{0.5cm}

In addition to the general formula above, we will need a certain special case. This is the situation when the domain of the parametrization $g$ happens to be a subspace of $\mathbb{R}^{n+m}$ and $g$ is itself is a graph. More precisely, for $z \in \mathbb{R}^{n+m}$, write it as $z = (z_{||}, z_\perp)$, where $z_{||} \in \mathbb{R}^n$ and $z_\perp \in \mathbb{R}^m$. Assume that $g$ is a function of $z_{||}$ and has the form

\ben
g(z_{||}) = \Big (z_{||}, g_\perp(z_{||}) \Big ).
\een

The derivative of $g$ is easily computed to be

\ben
Dg = \left ( \begin{array}{c} I \\ \hline Dg_\perp   \end{array}  \right ) ,
\een

where the right hand side stands for an $n+m$ by $n$ matrix with the top $n$ by $n$ chunk being the identity while the bottom $m$ by $n$ piece is the derivative of $g_\perp$. Hence, it follows from (\ref{eq:jacobianparam}) that 

\bes
\nonumber
Jg & = & \sqrt{ \det \Big (  (Dg)^T (Dg)        \Big )  } \\
\nonumber
& = & \sqrt{ \det \Big (  I + \big ( Dg_\perp \big )^T \big ( Dg_\perp   \big)    \Big )     } \\
\label{eqs:firstformula}
& = & \sqrt{ \det \Big (  I + \big (  Dg_\perp  \big) \big ( Dg_\perp    \big )^T   \Big )     },
\ees

where Sylvester's determinant theorem was used to get to the last line. \newline

Assume now that $h$ is a submersion, and write 

\ben
Dh = \left ( \begin{array}{ccc} D_{||}h & | & D_\perp h   \end{array} \right ),
\een

where $D_{||}h$ and $D_\perp h$ stands for the submatrices of $Dh$ containing the partial derivatives with respect to the components of $z_{||}$ and $z_\perp$ respectively. It follows from the implicit function theorem that

\ben
Dg_\perp = - (D_\perp h)^{-1} D_{||}h.
\een

Consequently, we get that

\bes
\nonumber
\det \Big ( (Dh) (Dh)^T \Big ) \det (D_\perp h)^{-2} & = & \det  \Big (  (D_\perp h)^{-1}  (Dh) (Dh)^T   \big ( (D_\perp h)^{-1}  \big )^T    \Big ) \\
\nonumber
& = & \det \Big (   \left ( \begin{array}{ccc} - Dg_\perp & | & I   \end{array} \right )    \left ( \begin{array}{c} - (Dg_\perp)^T \\ \hline I   \end{array}  \right )  \Big ) \\
\label{eqs:secondformula}
& = & \det \Big (  I + (Dg_\perp) (Dg_\perp)^T  \Big ).
\ees

Recalling (\ref{eq:jacobianlevel}), and putting together (\ref{eqs:firstformula}) and (\ref{eqs:secondformula}), we get that in the case we are describing

\ben
\frac{Jh}{Jg} = \pm \det \big ( D_\perp h   \big ).
\een

Plugging this into (\ref{eq:relation}), we get that

\be
\label{eq:relation2}
\int_{R^n} F(g(z_{||})) d z_{||} \simeq \int_{\mathbb{R}^{n+m}} F(z) \delta \big ( h(z) \big ) Jh_\perp dz,
\ee

where $Jh_\perp$ has the following two equivalent definitions

\bes
\nonumber
Jh_\perp & = &   \det \big ( D_\perp h   \big ) \\
\nonumber
& =  &   \txt{volume of the parallelepiped spanned by } Dh(v_1), \dots, Dh(v_m),\\
\label{eqs:jacobian3}
& &
\ees

where $v_1, \dots, v_m$ is an orthonormal basis of the space of $z_\perp$'s.\newline

So far, we know that (\ref{eq:relation2}) holds only if $h$ is a submersion. However, if $h$ is merely of constant rank, then proceeding as before by replacing $h$ with $\tilde{h} = O h$, where $O$ is an appropriately chosen rotation, and repeating the same arguments as before, we see that the relation continues to hold in that case as well. Of course, in this case, only the second definition of $Jh_\perp$ above makes sense where `volume' now means, as before, the $m$-dimensional Hausdorff measure.

\subsection*{Principal Chiral Model}

Let us move on now and consider the two dimensional $SU(N)$ principal chiral model. The fundamental fields in this case are group-valued maps $\phi: \mathbb{R}^2 \to SU(N)$ and the action is $\int_{\mathbb{R}^2} \txt{Tr} \Big ( (\phi^{-1} \partial_\mu \phi) (\phi^{-1} \partial^\mu \phi) \Big )$. It is natural to change variables to $A_\mu = \phi^{-1} \partial_\mu \phi$. Of course, this is exactly the form that a flat connection should have. Therefore, we shall attempt to rewrite the path integral of the model as an integral of a quadratic action, simply the Proca mass term, over the space of flat connections. \newline

We would like to use (\ref{eq:relation}) above assuming that it generalizes formally to the infinite dimensional setting. Our left hand side now is

\be
\label{eq:lhs}
\int \mathcal{D}\phi \,  e^{-\int_{\mathbb{R}^2} \txt{Tr} \Big ( (\phi^{-1} \partial_\mu \phi) (\phi^{-1} \partial^\mu \phi) \Big )   }   .
\ee

Therefore, we see that the analogue of the manifold $M$ is the space of fundamental fields. We claim that this is a Riemannian manifold and the formal measure $\mathcal{D} \phi$ is indeed the one induced from the Riemannian structure. To see this, note that a tangent vector can be represented by the derivative at $t = 0 $ of a curve of the form 

\be
\label{eq:curve}
t \to \Big (x \to \psi(x,t) \phi(x) \Big ),
\ee

where $\psi(x,0) = \mathbb{I}$. The inner product of two such vectors, corresponding to $\psi_1$ and $\psi_2$ is given by $\int_{\mathbb{R}^2} \txt{Tr} (\dot{\psi}_1(x,0) \dot{\psi}_2(x,0)   ) dx$. In other words, this is just the usual Riemannian metric on $SU(N)$ `summed' over all points of spacetime. Since the measure $\mathcal{D} \phi$ is similarly a product of Haar measures over all points of spacetime, and since the Haar measure is induced by the Riemannian structure on the group, we have our claim. \newline

Continuing the comparison of (\ref{eq:lhs}) with the left hand side of (\ref{eq:relation}) we see that the map $x \to g(x)$ corresponds to $\phi \to A_\mu = \phi^{-1} \partial_\mu \phi$ and that the analogue of $F(\cdot)$ is $e^{ - \int_{\mathbb{R}^2} \txt{Tr} \big ( A_\mu A^\mu  \big )}$.\newline

Let us now discuss the analogue of the right hand side of (\ref{eq:relation}) which corresponds to (\ref{eq:lhs}). Clearly, the analogue of $\mathbb{R}^{n+m}$ is the space of all $su(N)$-valued 1-forms. The analogue of the standard Euclidean metric on this space is 

\be
\label{eq:metric}
A_\mu \cdot \tilde{A}^\mu = \int_{\mathbb{R}^2} \txt{Tr}(A_\mu \tilde{A}^\mu) .
\ee

The analogue of the space $\mathbb{R}^l$ is, as we shall see below, the space of $su(N)$-valued 2-forms. The analogue of the standard Euclidean metric on this space is

\be
\label{eq:metric1}
C_{\mu \nu} \cdot \tilde{C}^{\mu \nu} =  \int_{\mathbb{R}^2} \txt{Tr}(C_{\mu \nu} \tilde{C}^{\mu \nu}).
\ee

Finally, the analogue of $dz$ would be $\mathcal{D}A$ which is formally equal to a product over Lebesgue measures at each point of spacetime. It remains to identify the analogues of $h$, $Jg$ and $Jh$. \newline

We shall begin with $Jg$. Let us thus compute the action of the differential of the map $\phi \to A_\mu$ on a tangent vector. Therefore, take a curve of the form (\ref{eq:curve}). Note that

\besn
\frac{d}{dt} \Big ( \phi^{-1} \psi^{-1} \partial_\mu( \psi \phi)  \Big ) \bigg |_{t = 0} & = & \phi^{-1}  \dot{ \Big ( \psi^{-1} } \Big) \partial_\mu( \psi \phi) \bigg |_{t=0} + \phi^{-1} \psi^{-1} \partial_\mu \Big ( \dot{\psi}  \phi  \Big ) \bigg |_{t=0} \\
& = & \phi^{-1} \partial_\mu \Big ( \dot{ \psi} \Big |_{t=0}  \Big ) \phi.
\eesn

We therefore see that the differential of our map is the composition of $\partial_\mu$ followed by the adjoint action.\footnote{Incidentally, this shows that the map $\phi \to A_\mu$ is an immersion, provided one imposes the Dirichlet boundary conditions (i.e. $\phi = \txt{identity}$) on the fields before removing the infrared regulator. The need for a careful consideration of the boundary conditions in the regularization is discussed at the beginning of chapter 2 of \cite{polyakovbook}.} Now, since the metric (\ref{eq:metric}) is invariant under the adjoint action, and since $\partial_\mu$ is independent of $\phi$, we see, recalling (\ref{eq:jacobian}), that the analogue of $Jg$ would be a constant. We can therefore discard it. \newline

Since $A_\mu = \phi^{-1} \partial_\mu \phi$, it follows, as mentioned at the beginning of this subsection, that $A_\mu$ is a flat connection. We thus take for $h$ the flatness constraint

\ben
G_{\mu \nu}(A) = \partial_\mu A_\nu - \partial_\nu A_\mu + [ A_\mu, A_\nu] = 0.
\een

It remains to find $Jh$. The derivative of $G$ is easily computed to be

\ben
G' \Big |_A v = \partial_{[\nu} v_{\mu]} + [ A_{[\mu}, v_{\nu]}],
\een

where as is traditional, the square brackets on the indices stand for antisymmetrization. Now, note that 

\besn
\phi^{-1} \bigg ( \partial_{\mu} \Big ( \phi v_\nu \phi^{-1}    \Big )    \bigg ) \phi & = & \phi^{-1} \bigg ( \phi \Big(\partial_\mu v_\nu \Big ) \phi^{-1}  + \Big (  \partial_{\mu} \phi \Big ) v_\nu \phi^{-1} + \phi v_\nu \Big (  \partial_\mu \phi^{-1} \Big ) \bigg ) \phi \\
& = & \partial_\mu v_\nu + \phi^{-1}( \partial_\mu \phi ) v_\nu - v_\nu \phi^{-1} ( \partial_\mu \phi),
\eesn

and thus

\ben
G' \Big |_A v = \phi^{-1} \bigg ( \partial_{[ \mu} \Big ( \phi v_{\nu]} \phi^{-1}    \Big )    \bigg ) \phi.
\een

We thus see that the derivative of $G$ consists of the composition of the adjoint action on the $su(N)$ one-form, followed by the exterior derivative $d$, followed by an adjoint action on the $su(N)$ two-form. Clearly, the adjoint action does not affect the metrics (\ref{eq:metric}) and (\ref{eq:metric1}). Recalling the definition of $Jh$ given in (\ref{eq:jacobian1}), we see that its analogue in this case coincides with the Jacobian of the exterior derivative. But this is again a constant independent of the field $A$ and thus can be discarded as well. \newline

Putting everything together, we have thus arrived at 

\ben
\int  e^{-\int_{\mathbb{R}^2} \txt{Tr} \Big ( (\phi^{-1} \partial_\mu \phi) (\phi^{-1} \partial^\mu \phi) \Big )   }   \mathcal{D}\phi   \simeq \int e^{- \int_{\mathbb{R}^2} \txt{Tr} (A_\mu A^\mu)} \delta \Big (G_{\mu\nu}(A) \Big ) \mathcal{D} A_\mu.
\een

To reduce clutter, we've only displayed the relation between the two partition functions since the inclusion of sources is trivial.

\subsection*{Yang-Mills}

We are ready to rewrite the Yang-Mills path integral

\ben
\int \mathcal{D} A e^{- S_{YM}(A)},
\een

where, we remind the reader, we are using the radial gauge for our gauge-fixing and thus $\mathcal{D}A$ stands for integration over the set of connections in this gauge. Note that the Faddeev-Popov determinant in this case is ignorable. \newline

Now, decompose the space of $\mathcal{B}$'s into the kernel of the map $\mathcal{T}$, denoting it by $\mathcal{B}_\perp$, and its orthogonal\footnote{Orthogonality is defined with respect to the metric (\ref{eq:metric}) suitably generalized (we integrate over the space of loops instead of $\mathbb{R}^2$).} complement, denoted by $\mathcal{B}_{||}$. Note that for a connection $A$ in the radial gauge, we have that $\mathcal{T}(\mathcal{B}(A)) = A$. It follows from this that the space of connections in the radial gauge can be identified with $\mathcal{B}_{||}$, and thus, the map $A \to \mathcal{B}(A)$ is in fact of the form $\mathcal{B}_{||} \to (\mathcal{B}_{||}, \mathcal{B}_\perp(\mathcal{B}_{||}))$. We can therefore use (\ref{eq:relation2}) and rewrite the Yang-Mills path integral as

\ben
\int \mathcal{D} \mathcal{B} e^{- S_{YM}(\mathcal{B})} \delta \big ( \mathcal{G}(\mathcal{B})   \big )  J\mathcal{G}_\perp.  
\een

We have already computed $D \mathcal{G}$ in the subsection dealing with the principal chiral model and have found it to be equal to the composition of an adjoint action with an exterior derivative followed by another adjoint action.\footnote{Of course, we've only done so for connections on $\mathbb{R}^2$, but the formal computations are the same in any dimension and thus, we assume, hold for functionals as well.} Now, note that  it follows from (\ref{eq:tmap}) that

\ben
\mathcal{T}\Big ( \phi(\sigma)  \mathcal{B}(\sigma) \phi^{-1}(\sigma) \Big ) = \phi(\sigma_x) \mathcal{T}(\mathcal{B})(x) (\phi(\sigma_x))^{-1},
\een

where $\phi(\sigma)$ is an $SU(N)$valued function of the loop $\sigma$. From this equation, it follows at once that the adjoint action takes $\mathcal{B}_\perp$ and $\mathcal{B}_{||}$ to themselves. Therefore, an orthonogmal basis of $\mathcal{B}_{||}$ is taken to an orthonormal basis. Combining this fact with the definition (\ref{eqs:jacobian3}), we get that $J\mathcal{G}_\perp$ is again a constant and thus can be discarded.\newline

Finally, putting together the result of this subsection with that of the previous one\footnote{More precisely, with its generalization to an infinite dimensional setting as opposed to $\mathbb{R}^2$.}, we see that we can write

\ben
\int \mathcal{D} A e^{- S_{YM}(A)} \simeq \int \mathcal{D} \mathcal{B} e^{- S_{YM}(\mathcal{B})} \delta \big ( \mathcal{G}(\mathcal{B})   \big )  \simeq \int \mathcal{D} \phi e^{- S_{YM}(\phi)},
\een

where $S_{YM}(A)$ is the usual Yang-Mills action in terms of a connection, $S_{YM}(\mathcal{B})$ is given by (\ref{eq:action}) and $S_{YM}(\phi)$ is the analogue of the principal chiral model action given by

\ben
S_{YM}(\phi) = \frac{D}{\mathcal{V}} \int_{(0,1)} ds \int \mathcal{D} \gamma \frac{  \txt{Tr}   \Big ( \phi^{-1}(\gamma) \frac{\delta \phi(\gamma)}{\delta \gamma^{\mu}(s)}    \phi^{-1}(\gamma) \frac{\delta \phi(\gamma)}{\delta \gamma_{\mu}(s)}   \Big )   }{ \dot{\gamma}_\nu(s) \dot{\gamma}^\nu(s)   }.
\een

Also, $\mathcal{D}{\mathcal{B}}$ stands for integration over the set of all transverse, nonanticipating 1-forms over the space of loops and $\mathcal{D}\phi$ stands for integration over the space of all $SU(N)$ valued maps on the space of loops.\newline

Again, we've only given the equality between the the partition functions since inclusion of sources is straightforward. \newline

We have thus achieved our goal of rewriting the Yang-Mills path integral, and have done so in two ways. Moreover, both of them are of the form of a Gaussian integral satisfying a quadratic constraint. \newline

\textbf{Acknowledgments:} The author would like to thank J. Merhej for reading a preliminary version of this paper and for the numerous comments which have greatly improved its readability.

\texttt{{\footnotesize Department of Mathematics, American University of Beirut, Beirut, Lebanon.}
}\\ \texttt{\footnotesize{Email address}} : \textbf{\footnotesize{tamer.tlas@aub.edu.lb}}


\begin{thebibliography}{99}
\bibitem{makeenko} Y. Makeenko, ``Methods of contemporary gauge theory'', CUP, Cambridge, (2002).
\bibitem{giles} R. Giles, ``Reconstruction of gauge potentials from Wilson loops'', Phys. Rev. D, (3), 24, no.8, 2160--2168, (1981).
\bibitem{mandelstam} S. Mandelstam, ``Quantum electrodynamics without potentials'', Ann. Physics, 19, 1--24, (1962).
\bibitem{mandelstam2} S. Mandelstam, ``Quantization of the gravitational field'', Proc. Roy. Soc. London Ser. A 270, 346--363, (1962).
\bibitem{mandelstam3} S. Mandelstam, ``Feynman rules for electromagnetic and Yang-Mills fields form the gauge independent field theoretic formalism'', Phys. Rev. 175, 1580--1623, (1968).
\bibitem{birula} I. Bia\l ynicki-Birula, ``Gauge-invariant Variables in the Yang-Mills Theory'', Bulletin de L'academic Polonaise des Sciences, Vol. XI, No. 3, 135--138, (1963).
\bibitem{polyakov} A. Polyakov, ``Gauge fields as rings of glue'', Nuclear Phys. B 164, no. 1, 171–188, (1980). 
\bibitem{hong} H.-M. Chan, P. Scharbach, S. T. Tsou, ``On loop space formulation of gauge theories'', Ann. Physics 166, no. 2, 396--421, (1986).
\bibitem{hongbook} H.-M. Chan, S. T. Tsou, ``Some elementary gauge theory concepts'', World Scientific Publishing Co., NJ, 1993.
\bibitem{gross} L. Gross, ``A Poincar\'{e} lemma for connection forms'', J. Funct. Anal. \textbf{63}, no.1, 1--46, (1985).
\bibitem{gross2} L. Gross, ``The Maxwell equations for Yang-Mills theory'', \textit{Mathematical quantum field theory and related topics}, 193--203, AMS, (1988).
\bibitem{driver} B. Driver, ``Classifications of bundle connection pairs by parallel translation and lassos'', J. Funct. Anal., 83, no.1, (1989).
\bibitem{polyakovbook} A. Polyakov, ``Gauge fields and strings'', Harwood Academic Publishers, (1987).
\bibitem{evans} L. Evans, R. Gariepy, ``Measure theory and fine properties of functions'', Textbooks in Mathematics, CRC Press, Boca Raton, (2015).
\bibitem{krantz} S. Krantz, H. Parks, ``Geometric integration theory'', Cornerstones, Birkh\"{a}user Boston, Inc., Boston, 2008.
\bibitem{construction} T. Tlas, ``On the Construction of Euclidean Invariant and Reflection Positive Measures on the Cylindrical Compactification of Distributions'', [arXiv: 2206.08131].
\end{thebibliography}
\end{document}